\documentclass[letterpaper,twocolumn,%showpacs,
prb
]{revtex4}

\usepackage{graphicx}
\usepackage{amssymb}
\usepackage{amsmath,amsfonts,latexsym}
\usepackage{array,tabularx,color}
\usepackage{dcolumn}               % Align table columns on decimal point
\usepackage[normalem]{ulem}
\usepackage{comment}

\newcommand{\vect}[1]{{\mathbf #1}}
\newcommand{\Frac}[2]{\displaystyle\frac{#1}{#2}}

\begin{document}

%\linenumbers

\title{Phase separation and collapse in Bose-Fermi mixtures with a
Feshbach resonance}

\author{Francesca M. Marchetti}
\email{fmm25@cam.ac.uk} %
\affiliation{Rudolf Peirels Centre for Theoretical Physics,
University of Oxford, 1 Keble Road, Oxford OX1 3NP, UK}

\author{Charles J. M. Mathy}
%\email{cmathy@Princeton.EDU}
\affiliation{Department of Physics, Princeton University, Princeton,
New Jersey 08544, USA}

\author{Meera M. Parish}
\email{mparish@princeton.edu} %
\affiliation{Department of Physics, Princeton University, Princeton,
New Jersey 08544, USA} %
\affiliation{Princeton Center for Theoretical Science, Princeton
University, Princeton, New Jersey 08544, USA}

\author{David A. Huse}
%\email{huse@princeton.edu}
\affiliation{Department of Physics, Princeton University, Princeton,
New Jersey 08544, USA}

\date{\today}
%\date{March 23, 2007}       % to fix in the last version!

\begin{abstract}
  We consider a mixture of single-component bosonic and fermionic
  atoms with an interspecies interaction that is varied using a
  Feshbach resonance. By performing a mean-field analysis of a
  two-channel model, which describes both narrow and broad Feshbach
  resonances, we find an unexpectedly rich phase diagram at zero
  temperature: Bose-condensed and non-Bose-condensed phases form a
  variety of phase-separated states that are accompanied by both
  critical and tricritical points. We discuss the implications of our
  results for the experimentally observed collapse of Bose-Fermi
  mixtures on the attractive side of the Feshbach resonance, and we
  make predictions for future experiments on Bose-Fermi mixtures
  close to a Feshbach resonance.
\end{abstract}


%03.75.Hh   Static properties of condensates: thermodynamical, statistical
%           and structural properties.
%03.75.Ss   Degenerate Fermi gases
%05.30.Fk   Fermion systems and electron gas

%71.10.Ca   Electron gas, Fermi gas

\maketitle

%%%%%%%%%%%%%%%%%%%%%%%%%%%%%%%%%%%%%%%
%%%%%%%%%%%%%%%%%%%%%%%%%%%%%%%%%%%%%%%
\section{Introduction}
The possibility of controlling the inter-atomic interaction strength
via Feshbach resonances has recently played a pivotal role in
investigating condensation phenomena in ultracold alkali gases. A
prominent example is the realization of condensed pairs in
two-component Fermi gases, where one has a crossover from the
weakly-interacting BCS regime to a Bose-Einstein condensate (BEC) of
diatomic molecules.~\cite{regal2004,zwierlein2004} By tuning the two
spin populations to be unequal, one can further explore quantum
phase transitions and phase separation (see, e.g.,
Refs.~\onlinecite{shin2006}, \onlinecite{partridge_prl06} and
references therein). Even richer scenarios are expected for
heteronuclear resonances in Bose-Fermi mixtures, because one can in
principle destroy the BEC by binding bosons and fermions into
\textit{fermionic} molecules.~\cite{powell2005} Moreover, a host of
novel phenomena has been predicted, such as spatial separation of
bosons and fermions induced by interspecies repulsive
interactions,~\cite{molmer_98,viverit_00} boson-mediated Cooper
pairing,~\cite{heiselberg_00,suzuki2008,kalas2008} density wave
phases in optical lattices,~\cite{lewenstein_04} and the formation
of polar molecules with dipolar interactions.

Following the recent detection of Feshbach resonances in Bose-Fermi
mixtures,~\cite{stan_04,inouye_04,ferlaino_06,deh2008} experiments
have begun to explore how the behavior of the $^{87}$Rb-$^{40}$K
system depends on the interspecies interaction
strength.~\cite{ospelkaus_06_3,ospelkaus_06,zaccanti_06,modugno_07,zirbel07}
Thus far, repulsive interactions (generated by approaching the
resonance from the side of positive scattering length $a_{BF}$) have
been observed to reduce the spatial overlap between fermions and the
BEC.
By contrast, attractive interactions can produce a sudden loss of
atoms, which has been attributed to enhanced three-body recombination
processes resulting from an unconstrained increase in the density.
Current mean-field
theories~\cite{molmer_98,modugno03,chui04,adhikari2004} predict that
this \emph{total} collapse of the mixture occurs above a critical
density or interaction strength, where the mixture is dynamically
unstable. However, these theories neglect any pairing between bosons
and fermions, which we show plays a crucial role in the stability of
the system.

In this paper, we address the possibility of pairing in Bose-Fermi
mixtures by considering a two-channel model that explicitly includes
fermionic molecules as an extra species of particle.
Such a model encompasses both broad and narrow Feshbach resonances.
We then determine the zero-temperature phase diagram for the
Bose-Fermi mixture as a function of the interaction strengths and
particle densities within a mean-field approximation.
Sufficiently close to the resonance, we always find phase separation
between a mixed BEC phase (where the BEC coexists with fermions) and
either a pure BEC, mixed BEC, normal or vacuum phase.
To our knowledge, this provides the first example of phase
separation in a Bose-Fermi mixture with \textit{attractive}
interactions --- e.g., $^3$He-$^4$He mixtures~\cite{graf67} and
polarized Fermi gases~\cite{shin2006,partridge_prl06} require
effectively repulsive interactions.
We explain why our results are consistent with the collapse observed
in current experiments and we discuss the best conditions for
observing the predicted phase separation.

The paper is organized as follows: Section~\ref{sec:model} describes
the two-channel model of a Bose-Fermi mixture and how it is related to
the single-channel theory. In Sec.~\ref{sec:phase_diagram}, we derive
the zero temperature phase diagram of a homogeneous mixture.  In
Sec.~\ref{sec:impli} we discuss its connection with current and future
experiments and we then consider how the phase diagram will manifest
itself in a trapped gas in Sec.~\ref{sec:trap}.  We present our
conclusions in Sec.~\ref{sec:conclusion}.

%
%%%%%%%%%%%%%%%%%%%%%%%%%%%%%%%%%%%%%%%
%%%%%%%%%%%%%%%%%%%%%%%%%%%%%%%%%%%%%%%
\section{Two-channel model}\label{sec:model}
We consider the two-channel Hamiltonian,
\begin{multline}
  \hat{H}_{2c} = \sum_{\vect{k}} \left( \xi_{\vect{k}}^f
    f^{\dag}_{\vect{k}} f^{}_{\vect{k}} + \xi_{\vect{k}}^b
    b^{\dag}_{\vect{k}} b^{}_{\vect{k}} + \xi_{\vect{k}}^{\psi}
    \psi^{\dag}_{\vect{k}} \psi^{}_{\vect{k}} \right) \\
  + \frac{g}{\sqrt{V}} \sum_{\vect{k}, \vect{k}'} \left(
    \psi^{\dag}_{\vect{k} + \vect{k}'} f^{}_{\vect{k}}
    b^{}_{\vect{k}'} + \text{h.c.} \right)\\
  + \frac{U_{bg}}{V} \sum_{\vect{k}, \vect{k}', \vect{q}}
  b^{\dag}_{\vect{k}} f^{\dag}_{\vect{k}'} f^{}_{\vect{k}' + \vect{q}}
  b^{}_{\vect{k} - \vect{q}}\\
  +\frac{\lambda}{V} \sum_{\vect{k}, \vect{k}', \vect{q}}
  b^{\dag}_{\vect{k}} b^{\dag}_{\vect{k}'} b^{}_{\vect{k}' + \vect{q}}
  b^{}_{\vect{k} - \vect{q}} \; ,
\label{eq:hamil}
\end{multline}
which describes a mixture of bosonic $b$ and fermionic $f$ atoms
coupled to a \emph{fermionic} closed channel molecule $\psi$ via the
interaction $g$, where $V$ is the three-dimensional volume. Setting
$\hbar=1$, the kinetic terms are given by
\begin{align*}
  \xi^f_{\vect{k}} &= \Frac{\vect{k}^2}{2m_f}
  - \mu_f\\
  \xi^b_{\vect{k}} &= \Frac{\vect{k}^2}{2m_b} - \mu_b\\
  \xi^{\psi}_{\vect{k}} &= \Frac{\vect{k}^2}{2(m_b + m_f)} - \mu_f -
  \mu_b + \nu \; ,
\end{align*}
where $\nu$ is the detuning from resonance and $\mu_{b}$ ($\mu_f$) is
the chemical potential for the bosons (fermions). We also include the
boson-boson interaction $\lambda = 2\pi a_{bb}/m_b$ and the background
boson-fermion interaction $U_{bg} = 4\pi a_{bg}/m$, with $m = 2m_f m_b
/(m_f + m_b)$. Clearly, we always require $\lambda > 0$ if we want the
Bose gas to be stable.

A key energy scale in our model is the width of the resonance
\begin{equation}
  \gamma = \Frac{g^2}{8\pi} m^{3/2}\ , \ \text{with}
  \ g = \sqrt{\frac{4\pi a_{bg}\Delta \mu \Delta B}{m}} \; ,
\end{equation}
where $\Delta B$ is the absolute width of the resonance in terms of
the magnetic field and $\Delta \mu$ is the difference in magnetic
moments between the closed and open channels (which is of order the
Bohr magneton). When compared to the boson condensation temperature,
$k_B T_0 = 2\pi [n_b/g_{3/2} (1)]^{2/3}/m_b$ (with $g_{3/2} (1) \simeq
2.612$) and the Fermi energy, $k_B T_F = (6\pi^2 n_f)^{2/3}/2m_f$, the
width of the resonance defines both narrow, $\gamma^2/k_B (T_0 + T_F)
\ll 1$, and wide, $\gamma^2/k_B (T_0 + T_F) \gg 1$, Feshbach
resonances.

At zero temperature, the Hamiltonian~\eqref{eq:hamil} can be exactly
diagonalized in the mean-field approximation, $\langle
b^{}_{\vect{k}}\rangle = \delta_{\vect{k},0} \sqrt{V}\Phi$, implying
that the two Fermi species dispersions, $\xi^f$ and $\xi^\psi$, are
now hybridized by the presence of the condensate:
\begin{multline}
  \xi^{F,\Psi} = \frac{1}{2}(\xi^f + U_{bg}\Phi^2+\xi^\psi) \\
  \pm \frac{1}{2} \sqrt{(\xi^f+ U_{bg}\Phi^2-\xi^\psi)^2 + 4
    g^2\Phi^2} .
\end{multline}
This leads to the following expression for the grand-canonical free
energy density $\Omega_{2c}^{(0)} (\mu_f, \mu_b) = \min_{\Phi}
f_{2c}(\Phi; \mu_f, \mu_b)$:
\begin{multline}
  f_{2c}(\Phi; \mu_f, \mu_b) = \frac{1}{V}\sum_{\vect{k}} \left[
    \Theta(-\xi_{\vect{k}}^F) \xi_{\vect{k}}^F +
    \Theta(-\xi_{\vect{k}}^\Psi) \xi_{\vect{k}}^\Psi \right]\\
  - \mu_b \Phi^2 + \lambda \Phi^4 \; .
\label{eq:grand}
\end{multline}
When $U_{bg}<0$, the free energy is unbounded from below:
\begin{equation*}
  f_{2c} (\Phi \to \infty; \mu_f, \mu_b) \propto - \Phi^5 \; .
\end{equation*}
Therefore, a repulsive background interaction strength, $U_{bg}>0$, is
required in order to have a stable solution. Later on, we will neglect
$U_{bg}$ in our calculations because, as we discuss in
Sec.~\ref{sec:phase_diagram}, it will not affect the main features of
the phase diagram provided it is sufficiently small.

Since experiments are performed at fixed densities, the chemical
potentials $\mu_f$ and $\mu_b$ have to be determined from the total
number of fermionic and bosonic atom densities,
\begin{align}
  n_{f,b} & = -\frac{\partial \Omega^{(0)}}{\partial \mu_{f,b}} \; .
\end{align}
Note that $n_f$ ($n_b$) includes fermions (bosons) bound into
molecules.
Despite the simplicity of our approach, we emphasize that our
mean-field analysis will be quantitatively accurate in the case of
narrow Feshbach resonances,~\cite{andreev2004} because Gaussian
fluctuations beyond mean-field scale like $\gamma/\sqrt{k_B (T_0 +
  T_F)}$.
We also expect it to provide a qualitative description of the phase
diagram for a broad Feshbach resonance as is the case in two-component
Fermi gases --- the phase diagram for polarized Fermi gases is
qualitatively similar for broad and narrow Feshbach
resonances.~\cite{parish2007}

%%%%%%%%%%%%%%%%%%%%%%%%%%%%%%%%%%%%%%%
%%%%%%%%%%%%%%%%%%%%%%%%%%%%%%%%%%%%%%%
\subsection{Connection to single-channel model}
\label{sec:single}
Before tackling the phase diagram, it is instructive to understand how
our model connects with existing single-channel theories, where the
closed channel molecule $\psi$ is ignored and we have Hamiltonian
\begin{multline}
  \hat{H}_{1c} = \sum_{\vect{k}} \left( \xi_{\vect{k}}^f
    f^{\dag}_{\vect{k}} f^{}_{\vect{k}} + \xi_{\vect{k}}^b
    b^{\dag}_{\vect{k}} b^{}_{\vect{k}}  \right) \\
  + \frac{1}{V} \sum_{\vect{k}, \vect{k}', \vect{q}} \left(U_{BF}
    b^{\dag}_{\vect{k}} f^{\dag}_{\vect{k}'} f^{}_{\vect{k}' +
      \vect{q}} b^{}_{\vect{k} - \vect{q}}
    + \lambda b^{\dag}_{\vect{k}} b^{\dag}_{\vect{k}'} b^{}_{\vect{k}'
      + \vect{q}} b^{}_{\vect{k} - \vect{q}}\right) \; ,
\end{multline}
with the effective Bose-Fermi interaction $U_{BF} = 4\pi a_{BF}/m$.
Such a theory is expected to be appropriate for a broad Feshbach
resonance, where there is only a small admixture of the closed channel
molecule, which can therefore be neglected.
However, a mean-field analysis of the single-channel model focuses on
densities only and does not take account of pairing correlations.
Within this approximation, the system becomes linearly unstable
when:~\cite{viverit_00}
\begin{align}
  n_f^{1/3} \geq \frac{4}{3} \frac{(6 \pi^2)^{2/3}}{2m_f}
  \frac{\lambda}{U_{BF}^2} \; .
\label{eq:unstable}
\end{align}
For the case of attractive interactions $U_{BF} < 0$, one can easily
show that the single-channel free energy is not bounded from below
($f_{1c} (\Phi \to \infty; \mu_f, \mu_b) \propto - \Phi^5$). As a
consequence, any state at finite density is metastable and, when
Eq.~\eqref{eq:unstable} is satisfied, the system is unstable to a
total collapse.~\cite{chui04} On the other hand, when $U_{BF} > 0$,
the free energy is bounded and the instability is towards phase
separation.

Now, we expect the two-channel model~\eqref{eq:hamil} to map to a
single-channel Hamiltonian when we take the limit $\nu \to + \infty$,
$g \to + \infty$, while holding $-g^2/\nu \equiv U_{BF} <0$ fixed,
because then the closed channel molecule $\psi$ effectively disappears
from the problem while the scattering length $a_{BF}$ is kept fixed.
Indeed, we find that our two-channel mean-field
theory~\eqref{eq:grand} in the above limit is formally equivalent to
the single-channel mean-field theory on the \emph{attractive} side of
the resonance with $U_{BF} < 0$.
Thus, in the low density regime $(T_0 + T_F) \ll \gamma^2/k_B$ we
expect the two-channel theory to reduce to the single-channel theory.
Therefore, we expect to have linear instabilities and first order
transitions in the low-density region of the phase diagram.
In addition, we do not expect our mean-field theory to include the
pairing instabilities that have been shown to exist in the
single-channel model,~\cite{kagan04,storozhenko05} although we
speculate that the narrow Feshbach resonance limit may capture the
qualitative features of these pairing correlations.

\begin{figure}
\begin{center}
\includegraphics[width=0.48\textwidth,angle=0]{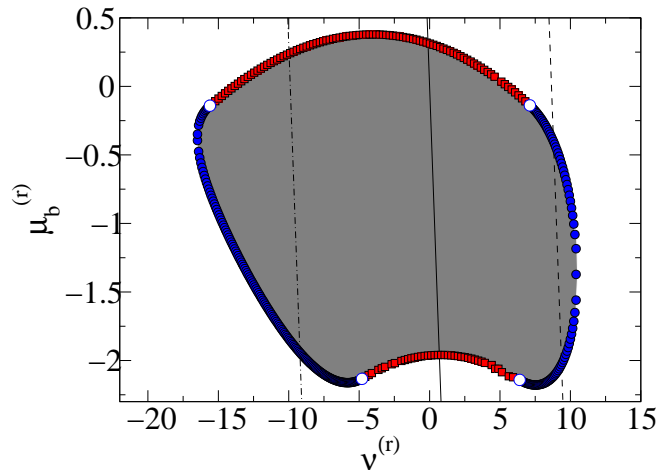}
\end{center}
\caption{(Color online). Surface of first-order phase transitions in
  the 3D space $\{\mu_f^{(r)},\nu^{(r)},\mu_b^{(r)}\} \equiv
  c^{1/3}\{c^{1/3} \mu_f/\gamma^2 , c^{1/3}(\nu - \mu_b)/\gamma^2 ,
  \mu_b/\gamma^2\}$, where $c = \lambda m_b^{2/3} \gamma$, that has
  been projected onto the $(\nu^{(r)},\mu_b^{(r)})$ plane.  The gray
  shaded area corresponds to the first-order region. For $|\nu^{(r)}|$
  greater than the tetracritical points $O(\Phi^2) = O(\Phi^4)=
  O(\Phi^6) = 0$ (open circles), we have lines of tricritical points
  represented by the (blue) filled circles. Otherwise, they are
  replaced by critical points [(red) filled squares].  The straight
  lines correspond to the cuts at fixed $\nu/\gamma^2$ and $c$ in
  Fig.~\ref{fig:figu1}, where we have $c\simeq 0.044$ and
  $\nu/\gamma^2 =$ -80 (dotted-dashed), 0 (solid) and 70 (dashed),
  going from left to right.}
\label{fig:figu2}
\end{figure}
\begin{figure*}
\begin{center}
\includegraphics[width=1\linewidth,angle=0]{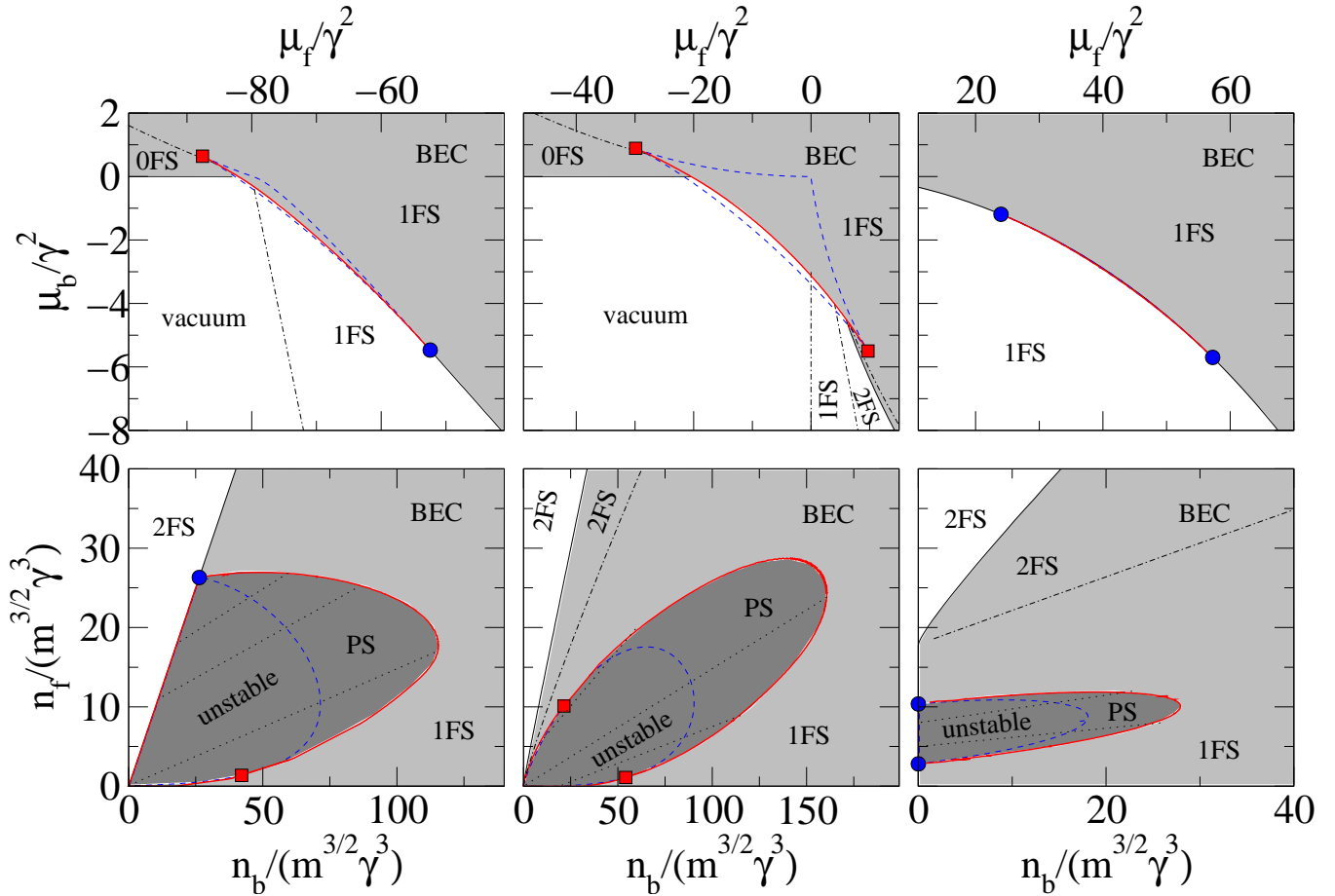}
\end{center}
\caption{(Color online) Zero temperature, mean-field phase diagrams
  for $^{23}$Na-$^{6}$Li mixtures, where $c=\lambda m_b^{3/2} \gamma
  \simeq 0.044$. The top and bottom rows correspond to chemical
  potential and density space, respectively, while the columns
  represent different detunings: $\nu/\gamma^2=-80$ ($a_{BF} \simeq
  -475 a_0$), $\nu/\gamma^2=0$, and $\nu/\gamma^2=70$ ($a_{BF} \simeq
  543 a_0$), going from left to right. The various phases can be
  distinguished based on the number of Fermi surfaces (FS) and whether
  or not there is a BEC (light gray shaded region).  The solid thick
  (red) lines represent first-order phase transitions, which are
  accompanied by regions of phase separation (PS --- dark gray shaded
  region) in density space, while the thin solid (black) lines of the
  phase boundaries are continuous. The dotted-dashed lines separate
  regions with a different number of Fermi surfaces.  The dotted lines
  that connect points on the first-order boundary depict which phases
  constitute the phase-separated state at a given density. A filled
  (blue) circle denotes a tricritical point, a filled (red) square
  denotes a critical point, and a dashed (blue) line denotes a
  spinodal line.}
\label{fig:figu1}
\end{figure*}
%

%%%%%%%%%%%%%%%%%%%%%%%%%%%%%%%%%%%%%%%
%%%%%%%%%%%%%%%%%%%%%%%%%%%%%%%%%%%%%%%
\section{Phase Diagram}
\label{sec:phase_diagram}
In the following, we will take the mass ratio to be $m_f/m_b \simeq
0.26$ as in $^{23}$Na-$^{6}$Li mixtures, because this atomic system
has a narrow Feshbach resonance~\cite{stan_04,gacesa2008} and a small
repulsive $U_{bg}$. Thus we are likely to observe experimentally the
phase-separated states we predict here. While the behavior of
$^{23}$Na-$^6$Li mixtures is yet to be explored in detail,
inter-species Feshbach resonances have already been identified
experimentally.
We emphasize that changing the mass ratio will only affect our results
qualitatively and so our phase diagram will also be applicable to
other atomic systems.

If one focuses on pairing only, it is clear that at the fixed density
$n_f \geq n_b$ there is a transition as we cross the resonance: the
BEC phase ($\Phi \ne 0$) will be completely depleted ($\Phi = 0$) by
the binding of bosons into fermionic molecules. (Note that the
formation of molecules can also involve a phase transition where the
number of Fermi surfaces changes).
This transition was shown to be continuous for the limiting case of
vanishing coupling $g=0$.~\cite{yabu2003} Assuming the transition
remains second order for $g \ne 0$, as in
Ref.~\onlinecite{powell2005}, the phase boundaries are found by
solving $O(\Phi^2) \equiv \partial f_{2c}/\partial \Phi^2|_{\Phi=0} =
0$.
However, as we anticipated in Sec.~\ref{sec:single}, in general we
find that there can also be first order transitions between two BEC
phases as well as between BEC and normal (N) phases. Before we explain
the main features of the phase diagram, it is useful to examine how
the second order transition becomes first order.

Precursors of first order transitions can be found in tricritical
points, where $O(\Phi^2)=0$ and $O(\Phi^4) \equiv \partial^2
f_{2c}/\partial(\Phi^2)^2|_{\Phi=0} = 0$. Surprisingly, the mean-field
energy $f_{2c}(\Phi;\mu_f,\mu_b)$ can be expressed in terms of just
three independent dimensionless parameters, and thus the phase
transitions can be completely characterized by the dimensionless
parameters
\begin{equation}
  \{\mu_f^{(r)}, \nu^{(r)}, \mu_b^{(r)}\} =
  c^{1/3} \{c^{1/3}\frac{\mu_f}{\gamma^2}, c^{1/3}\frac{\nu -
    \mu_b}{\gamma^2} ,
  \frac{\mu_b}{\gamma^2}\} \; ,
\end{equation}
where $c = \lambda m_b^{2/3} \gamma$. Referring to
Fig.~\ref{fig:figu2}, in these units we find that for $|\nu^{(r)}|$
greater than the tetracritical values at which $O(\Phi^2) = O(\Phi^4)=
O(\Phi^6) = 0$ (open circles), there are lines of tricritical points
[(blue) filled circles] where the transition changes from second to
first order. Beyond the tetracritical points, the tricritical points
are replaced by critical points [(red) filled squares]: At a critical
point two equal energy minima of $f_{2c}(\Phi; \mu_f, \mu_b)$ merge.
In this case, an expansion in $\Phi^2$ cannot be exploited any longer.
However, this failure of the expansion is not so surprising because,
when $\lambda = 0$, the mean-field potential~\eqref{eq:grand} is
unbounded from below ($f_{2c}(\Phi \to \infty; \mu_f, \mu_b) \propto
-\Phi^{5/2}$). This non-analytic dependence on $\Phi$ comes from the
hybridized Fermi dispersions $\xi^{F,\Psi}$.

Therefore, to summarize, we find that the first-order regime is
confined by either tricritical or critical points within a region
about the origin of the $\{\mu_f^{(r)},\nu^{(r)},\mu_b^{(r)}\}$
parameter space. Outside of this region, we only find continuous
transitions. From the definition of
$\{\mu_f^{(r)},\nu^{(r)},\mu_b^{(r)}\}$, one can see that the ratio
$\lambda m_b^{3/2}/\gamma^2$ sets the energy scale of the problem and
determines the chemical potentials at which the tricritical points are
replaced by critical points in Fig.~\ref{fig:figu2}.
As explained above, one expects first order transitions to appear in
the limit $\lambda \to 0$.
However, note also that one can access the regime of first order
transitions even when the repulsion between bosons $\lambda$ is large,
by making $\mu_f$, $\nu$ and $\mu_b$ sufficiently small --- as a
consequence we expect these results to apply in the low density limit
close to resonance.
We have checked that a small repulsive background interaction $U_{bg}$
only brings small quantitative changes to this result by slightly
reducing the size of the first order region.

While one can parameterize the entire phase diagram using the rescaled
parameters $\{\mu_f^{(r)},\nu^{(r)},\mu_b^{(r)}\}$, from the point of
view of real systems, it makes more sense to consider the parameters
$\{c\equiv \lambda m_b^{3/2} \gamma, \nu/\gamma^2, \mu_b/\gamma^2,
\mu_f/\gamma^2\}$. For a given experiment, $c$ is fixed by the typical
boson-boson interaction strength and width of the resonance, while
$\nu/\gamma^2$ is determined by the Bose-Fermi scattering length
$a_{BF} = -2\gamma /\sqrt{m}\nu$.  Therefore, the phase diagram can be
plotted as a function of the chemical potentials (top row of
Fig.~\ref{fig:figu1}).  These slices correspond to planes in
Fig.~\ref{fig:figu2} given by $\mu_b^{(r)} = -\nu^{(r)}/c^{1/3} +
c^{1/3} \nu/\gamma^2$. In order to be relevant to experiments on
$^{23}$Na-$^{6}$Li mixtures, we use the scattering lengths $a_{bg} =
13.0a_0$ and $a_{bb} = 85a_0$, where $a_0$ is the Bohr radius, and
take the resonance width to be $\Delta B = 2.2$G.~\cite{gacesa2008}
For these values of the parameters, we have $c \simeq 0.044$ and
$a_{BF} = - 3.8 \times 10^4 a_0 \gamma^2/\nu$.  However, the
qualitative behavior of the phase diagram as depicted in
Fig.~\ref{fig:figu1} will apply for all $c \lesssim 25$. In this
regime, first order transitions appear only in the finite interval
around unitarity, $c^{2/3} \nu/\gamma^2 \in [-18,11]$. At the
resonance, we find two critical points (and two accompanying critical
endpoints, where first and second order transition meet), while moving
far from the resonance in either direction, first one and then both
critical points are replaced by tricritical points, until eventually
for large enough detuning the first order transition region disappears
altogether.

In terms of the phase diagram in density space (bottom row of
Fig.~\ref{fig:figu1}), first order transitions imply phase separation
(PS) between BEC and N (BEC and BEC) close to a tricritical (critical)
point. Here, N satisfies $n_f=n_b$ when $\nu<0$ and $n_b=0$ when
$\nu>0$.
Both BEC and N phases can be further characterized by the number of
Fermi surfaces (FS). In the N region, the second-order boundaries
between regions with different numbers of FS are defined by
$\mu_f=0$ ($n_b = n_f$) and $\mu_b=-\mu_f+\nu$ ($n_b=0$). In the BEC
phase instead one has to impose the condition $g^2 \Phi^2 = \mu_f
(\mu_f + \mu_b - \nu)$.
A special case of PS occurs when the N phase is the vacuum: Physically
this is equivalent to a \emph{partial} collapse of the system to
higher densities.
%We will return to this point later when we compare our results with
%experiment
On the attractive side of the resonance, $\nu>0$, and for small enough
densities, $(T_0 + T_F) \ll \gamma^2/k_B$, we expect to recover the
results of the single-channel theory. In particular, close enough to
the resonance and for small enough densities, the condition for linear
instability becomes independent of the boson density, resembling
Eq.~\eqref{eq:unstable}.

At this point, a question that naturally arises is: why does one
observe phase separation instead of the \emph{total} collapse
predicted by the single-channel mean-field theory? The answer is that
the presence of closed-channel fermionic molecules stabilizes the
large $\Phi$ behavior and constrains unbounded increases in density,
converting the total collapse into phase separation. One can
intuitively understand this as follows: once the Bose-Fermi mixture
becomes unstable, the resulting increase in density will likewise
increase the population of (virtual) closed-channel fermionic
molecules, which in turn will exert an increasing Fermi pressure.
Eventually, this Fermi pressure will balance the negative pressure of
the attractive interactions so that the gas becomes stable at a finite
density. This scenario is perhaps most clearly illustrated close to
unitarity, where phase separation takes the form of a partial
collapse.

Within the PS region it is possible to distinguish a metastable and an
unstable (or spinodal) region separated by a spinodal line, where
minimum and maximum of $f_{2c}(\Phi; \mu_f, \mu_b)$ merge. Inside the
spinodal region, the dynamics of phase separation following a sudden
quench proceeds via a linear instability (see, e.g.,
Ref.~\onlinecite{lamacraft07} and references therein), while it is
characterized by nucleation in the metastable region. A calculation of
the spinodal lines has also been carried out in
Ref.~\onlinecite{powell2005} at a fixed $n_b$, although the
possibility of phase separation was not considered.

In addition, Ref.~\onlinecite{powell2005} suggests that, when
$n_f<n_b$, there is a smooth crossover at fixed density from the
atomic to the molecular side of the resonance within the same BEC-1FS
phase. We find that this crossover exists for sufficiently large
densities, while phase separation intervenes at smaller densities $n_f
\lesssim 30 m^{3/2} \gamma^3$ and $n_b \lesssim 170 m^{3/2} \gamma^3$
(Fig.~\ref{fig:figu1}). More generally, the size of the
phase-separated region scales like $\gamma^2/\lambda$ for both $n_f$
and the molecular component of $n_b$, while the condensed component of
$n_b$ scales like $\gamma^2/(\lambda c^{1/3})$.

Finally, we note that the narrow Feshbach resonance regime,
$\gamma^2/k_B (T_0 + T_F) < 1$, corresponds to densities $n_f > 0.02
m^{3/2} \gamma^3$ or $n_b > 0.6 m^{3/2} \gamma^3$, therefore our
mean-field treatment should be reasonable for the region of interest.
Certainly, the unstable region is a robust feature of the overall
phase diagram, because we can arbitrarily increase the size of it in
density-space by decreasing $\lambda/\gamma^2$ or $\lambda
c^{1/3}/\gamma^2$.

\begin{figure}
\begin{center}
\includegraphics[width=0.8\linewidth,angle=0]{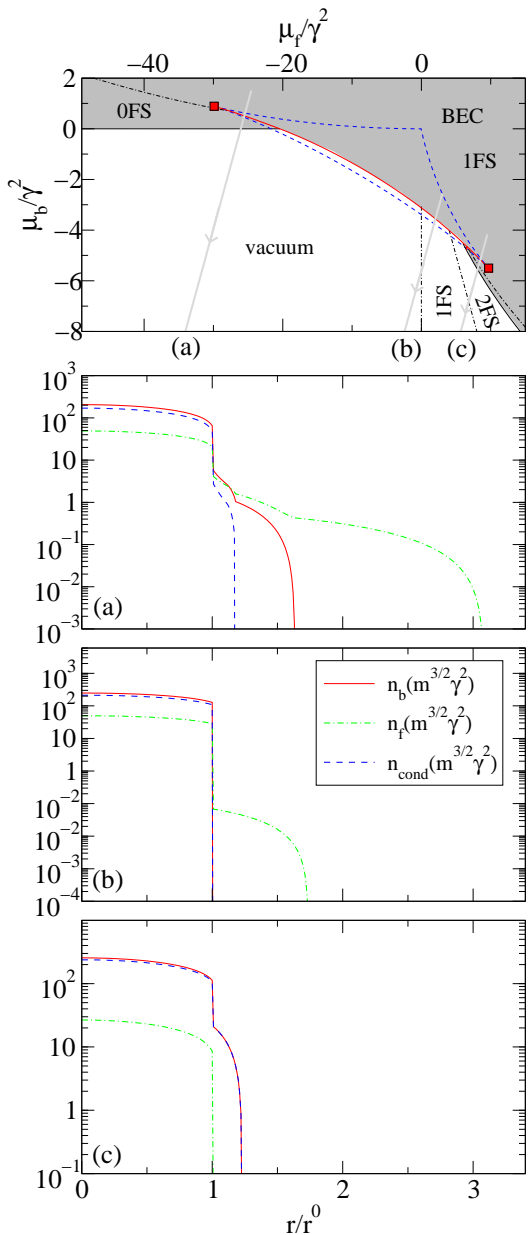}
\end{center}
\caption{(Color online). Trap density profiles at $\nu=0$
  for boson number $n_b$, fermion number $n_f$ and condensed boson
  number $n_{\text{cond}} \equiv \Phi^2/g^2$ in units of $m^{3/2}
  \gamma^3$, where $\lambda m_b^{3/2} \gamma \simeq 0.044$. The
  profiles correspond to three different cuts, (a), (b), and (c),
  across the chemical potential phase diagram as shown in the top
  panel, where the gray arrows represent trajectories from the center
  to the edge of the trap. In all three cases $r= r_0$ has been fixed
  as the point at which the first order transition occurs.}
\label{fig:figu3}
\end{figure}
%

%%%%%%%%%%%%%%%%%%%%%%%%%%%%%%%%%%%%%%%
%%%%%%%%%%%%%%%%%%%%%%%%%%%%%%%%%%%%%%%
\section{Implications for experiment}
\label{sec:impli}
The existence of three-body recombination in Bose-Fermi mixtures poses
a major challenge to investigating experimentally the phase diagram we
have described in Sec.~\ref{sec:phase_diagram}.
In current experiments on $^{87}$Rb-$^{40}$K mixtures, the behavior of
the mixture is dominated by the large attractive background
interaction --- one even achieves collapse far from the resonance by
increasing the density.~\cite{ospelkaus_06,ospelkaus_06_3,zaccanti_06}
Even if one ignores the background scattering length, we find that the
phase-separated states close to resonance, as in Fig.~\ref{fig:figu1},
evaluated for the parameters of $^{87}$Rb-$^{40}$K, involve such a
large increase in density (becoming of order $10^{16}$cm$^{-3}$ or
more) that they will be destroyed by three-body recombination. Thus,
we do not expect experiments on $^{87}$Rb-$^{40}$K mixtures to reveal
the rich variety of phase-separated states we have predicted. On the
other hand, a density of $10^{12}$cm$^{-3}$ in $^{23}$Na-$^6$Li
mixtures corresponds to $n_b/m^{3/2}\gamma^3 \sim 1$ and so the
phase-separated state possesses densities of order
$10^{13}-10^{14}$cm$^{-3}$, which should be easily accessible
experimentally.

The dominant three-body process involves 1 fermionic atom and 2
bosonic atoms, with recombination rate $\Gamma \propto a_{BF}^4 n_f
n_b^2$ away from resonance.~\cite{dincao2005} Thus, to minimize
$\Gamma$ in the phase-separated state, we must reduce the densities
and/or $|a_{BF}|$ at which phase separation first appears. Assuming
the bosons are mostly condensed and using the scalings described
previously, we get $\Gamma/n_b \propto m^{3/2 }\lambda^{1/3}
\gamma^{7/8}$ and $\Gamma/n_f \propto m \gamma^2$. Therefore, we must
consider small $\gamma$, such as the Feshbach resonance for
$^{23}$Na-$^6$Li mixtures,~\cite{stan_04} and perhaps small $\lambda$,
although the dependence of $\Gamma$ on $\lambda$ is sensitive to the
precise power of $|a_{BF}|$, which in turn can depend on the width of
the resonance.~\cite{petrov2004} One might also worry about three-body
losses resulting from collisions between the closed-channel molecule
$\psi$ and the open-channel atoms. However, we point out that the
`bare' molecule-atom interaction is generally unrelated to the
resonance-induced interaction in the open channel and is, thus,
generally negligible.  Therefore, any collisions between
closed-channel molecules and atoms will have to be due to effective
interactions resulting from higher-order processes that involve
multiple scattering among the atoms in the open-channel. Such
processes on the attractive side of the resonance should already be
encompassed by our analysis of three-body collisions above.

On the repulsive side of the resonance, $U_{BF} > 0$, the mean-field
single channel theory predicts phase separation between atomic bosons
and fermions induced by the repulsion.~\cite{viverit_00} One can
instead access the regime of model~\eqref{eq:hamil} by starting from
$a_{BF} <0$ and sweeping through the resonance to the molecular side,
$a_{BF} > 0$. Indeed, while most of the experiments have focused on
the repulsive Bose-Fermi interaction, fermionic molecules have been
very recently produced.~\cite{ospelkaus_06_2,modugno_07,zirbel07}
However, these studies currently paint a pessimistic picture of the
stability of these \emph{real} (as opposed to virtual closed-channel)
molecules with respect to three-body recombination.

%%%%%%%%%%%%%%%%%%%%%%%%%%%%%%%%%%%%%%%
%%%%%%%%%%%%%%%%%%%%%%%%%%%%%%%%%%%%%%%
\subsection{Trapped gases}
\label{sec:trap}
In principle, one can detect phase separation using \emph{in situ}
absorption measurements. To extract the behavior of the trapped gas
from the uniform phase diagram (Fig.~\ref{fig:figu1}), we use the
local density approximation to convert the effects of the trap into
spatially varying chemical potentials. As shown in
Fig.~\ref{fig:figu3}, the spatial trajectory in a trap is then
represented by the straight line $\mu_{f,b} (\vect{r})/\gamma^2 =
\mu_{f,b} (0)/\gamma^2 - V(\vect{r})/\gamma^2$ in the phase diagram,
assuming that the fermionic and bosonic atoms experience the same
harmonic trapping potential $V(\vect{r})$.~\cite{zaccanti_06}
Phase separation in a trap occurs when the first-order boundary
between BEC and N is crossed, yielding a discontinuity in the density
profiles. In this case, the central phase is always a BEC (with 1FS),
while the surrounding phases can be either BEC or N (with different
numbers of FS). In particular, the partially-collapsed gas will
exhibit a sharp interface with the vacuum. The presence of density
discontinuities will provide the primary signature for the phase
separation predicted here.

%%%%%%%%%%%%%%%%%%%%%%%%%%%%%%%%%%%%%%%
%%%%%%%%%%%%%%%%%%%%%%%%%%%%%%%%%%%%%%%
\section{Conclusion}
\label{sec:conclusion}
In conclusion, we have analyzed the zero temperature phase diagram
for a Bose-Fermi mixture with an interaction strength tunable via a
Feshbach resonance. By making use of a two-channel model, we have
found a finite region of phase separation around resonance that is
bounded by either tricritical or critical points. Close to
unitarity, phase separation takes the form of a \emph{partial}
collapse of the system, where phase separation occurs between a
higher density mixed BEC phase (BEC coexisting with either atomic or
molecular fermions) and the vacuum. Such a sudden increase in
density implies a larger three-body recombination, and thus our
phase-separated states may be a challenge to realize experimentally.
Indeed, current experiments on $^{87}$Rb-$^{40}$K mixtures are
dominated by total collapse in the case of attractive
interactions.~\cite{ospelkaus_06,ospelkaus_06_3,zaccanti_06}
However, we have argued that mixtures with a small resonance width
and a small repulsive background interaction, such as
$^{23}$Na-$^6$Li mixtures,~\cite{stan_04}, stand a better chance of
realizing the phase diagram we predict.

Finally, we note that in this work we have neglected the possibility
of fermionic superfluidity induced by density fluctuations of the
bosonic condensate: For a spin polarized Fermi gas, boson-mediated
$p$-wave~\cite{suzuki2008} and $s$-wave
odd-frequency~\cite{kalas2008} Cooper pairing have been recently
analyzed for the single-channel model. In both cases it has been
found that the conditions for Cooper pairing are favorable for a
repulsive enough Bose-Fermi interaction strength $U_{BF}$. However,
at least for $^{23}$Na-$^{6}$Li mixtures, we expect these phases to
occur at densities much larger than the ones considered in the phase
diagram of Fig.~\ref{fig:figu1}.

%%%%%%%%%%%%%%%%%%%%%%%%%%%%%%%%%%%%%%%%%%%%%%%%%
%%%%%%%%%%%%%%%%%%%%%%%%%%%%%%%%%%%%%%%%%%%%%%%%%
\acknowledgments We are grateful to J. Chalker, V. Gurarie, P. B.
Littlewood, and G. Modugno for stimulating discussions. FMM would
like to acknowledge the financial support of the EPSRC. This
research was supported in part by the National Science Foundation
under Grant Numbers PHY05-51164, DMR-0645461 and DMR-0213706.
%%%%%%%%%%%%%%%%%%%%%%%%%%%%%%%%%%%%%%%%%%%%%%%%%

%\bibliography{LettRefsBF}

\begin{thebibliography}{34}
\expandafter\ifx\csname
natexlab\endcsname\relax\def\natexlab#1{#1}\fi
\expandafter\ifx\csname bibnamefont\endcsname\relax
  \def\bibnamefont#1{#1}\fi
\expandafter\ifx\csname bibfnamefont\endcsname\relax
  \def\bibfnamefont#1{#1}\fi
\expandafter\ifx\csname citenamefont\endcsname\relax
  \def\citenamefont#1{#1}\fi
\expandafter\ifx\csname url\endcsname\relax
  \def\url#1{\texttt{#1}}\fi
\expandafter\ifx\csname urlprefix\endcsname\relax\def\urlprefix{URL
}\fi \providecommand{\bibinfo}[2]{#2}
\providecommand{\eprint}[2][]{\url{#2}}

\bibitem[{\citenamefont{Regal et~al.}(2004)\citenamefont{Regal, Greiner, and
  Jin}}]{regal2004}
\bibinfo{author}{\bibfnamefont{C.~A.} \bibnamefont{Regal}},
  \bibinfo{author}{\bibfnamefont{M.}~\bibnamefont{Greiner}}, \bibnamefont{and}
  \bibinfo{author}{\bibfnamefont{D.~S.} \bibnamefont{Jin}},
  \bibinfo{journal}{Phys. Rev. Lett.} \textbf{\bibinfo{volume}{92}},
  \bibinfo{pages}{040403} (\bibinfo{year}{2004}).

\bibitem[{\citenamefont{Zwierlein et~al.}(2004)\citenamefont{Zwierlein, Stan,
  Schunck, Raupach, Kerman, and Ketterle}}]{zwierlein2004}
\bibinfo{author}{\bibfnamefont{M.~W.} \bibnamefont{Zwierlein}},
  \bibinfo{author}{\bibfnamefont{C.~A.} \bibnamefont{Stan}},
  \bibinfo{author}{\bibfnamefont{C.~H.} \bibnamefont{Schunck}},
  \bibinfo{author}{\bibfnamefont{S.~M.~F.} \bibnamefont{Raupach}},
  \bibinfo{author}{\bibfnamefont{A.~J.} \bibnamefont{Kerman}},
  \bibnamefont{and} \bibinfo{author}{\bibfnamefont{W.}~\bibnamefont{Ketterle}},
  \bibinfo{journal}{Phys.\ Rev.\ Lett.} \textbf{\bibinfo{volume}{92}},
  \bibinfo{pages}{120403} (\bibinfo{year}{2004}).

\bibitem[{\citenamefont{Shin et~al.}(2006)\citenamefont{Shin, Zwierlein,
  Schunck, Schirotzek, and Ketterle}}]{shin2006}
\bibinfo{author}{\bibfnamefont{Y.}~\bibnamefont{Shin}},
  \bibinfo{author}{\bibfnamefont{M.~W.} \bibnamefont{Zwierlein}},
  \bibinfo{author}{\bibfnamefont{C.~H.} \bibnamefont{Schunck}},
  \bibinfo{author}{\bibfnamefont{A.}~\bibnamefont{Schirotzek}},
  \bibnamefont{and} \bibinfo{author}{\bibfnamefont{W.}~\bibnamefont{Ketterle}},
  \bibinfo{journal}{Phys. Rev. Lett.} \textbf{\bibinfo{volume}{97}},
  \bibinfo{pages}{030401} (\bibinfo{year}{2006}).

\bibitem[{\citenamefont{Partridge et~al.}(2006)\citenamefont{Partridge, Li,
  Liao, Hulet, Haque, and Stoof}}]{partridge_prl06}
\bibinfo{author}{\bibfnamefont{G.~B.} \bibnamefont{Partridge}},
  \bibinfo{author}{\bibfnamefont{W.}~\bibnamefont{Li}},
  \bibinfo{author}{\bibfnamefont{Y.~A.} \bibnamefont{Liao}},
  \bibinfo{author}{\bibfnamefont{R.~G.} \bibnamefont{Hulet}},
  \bibinfo{author}{\bibfnamefont{M.}~\bibnamefont{Haque}}, \bibnamefont{and}
  \bibinfo{author}{\bibfnamefont{H.~T.~C.} \bibnamefont{Stoof}},
  \bibinfo{journal}{Phys. Rev. Lett.} \textbf{\bibinfo{volume}{97}},
  \bibinfo{eid}{190407} (\bibinfo{year}{2006}).

\bibitem[{\citenamefont{Powell et~al.}(2005)\citenamefont{Powell, Sachdev, and
  B\"{u}chler}}]{powell2005}
\bibinfo{author}{\bibfnamefont{S.}~\bibnamefont{Powell}},
  \bibinfo{author}{\bibfnamefont{S.}~\bibnamefont{Sachdev}}, \bibnamefont{and}
  \bibinfo{author}{\bibfnamefont{H.~P.} \bibnamefont{B\"{u}chler}},
  \bibinfo{journal}{Phys. Rev. B} \textbf{\bibinfo{volume}{72}},
  \bibinfo{pages}{024534} (\bibinfo{year}{2005}).

\bibitem[{\citenamefont{M\o{}lmer}(1998)}]{molmer_98}
\bibinfo{author}{\bibfnamefont{K.}~\bibnamefont{M\o{}lmer}},
  \bibinfo{journal}{Phys. Rev. Lett.} \textbf{\bibinfo{volume}{80}},
  \bibinfo{pages}{1804} (\bibinfo{year}{1998}).

\bibitem[{\citenamefont{Viverit et~al.}(2000)\citenamefont{Viverit, Pethick,
  and Smith}}]{viverit_00}
\bibinfo{author}{\bibfnamefont{L.}~\bibnamefont{Viverit}},
  \bibinfo{author}{\bibfnamefont{C.~J.} \bibnamefont{Pethick}},
  \bibnamefont{and} \bibinfo{author}{\bibfnamefont{H.}~\bibnamefont{Smith}},
  \bibinfo{journal}{Phys. Rev. A} \textbf{\bibinfo{volume}{61}},
  \bibinfo{pages}{053605} (\bibinfo{year}{2000}).

\bibitem[{\citenamefont{Heiselberg et~al.}(2000)\citenamefont{Heiselberg,
  Pethick, Smith, and Viverit}}]{heiselberg_00}
\bibinfo{author}{\bibfnamefont{H.}~\bibnamefont{Heiselberg}},
  \bibinfo{author}{\bibfnamefont{C.~J.} \bibnamefont{Pethick}},
  \bibinfo{author}{\bibfnamefont{H.}~\bibnamefont{Smith}}, \bibnamefont{and}
  \bibinfo{author}{\bibfnamefont{L.}~\bibnamefont{Viverit}},
  \bibinfo{journal}{Phys. Rev. Lett.} \textbf{\bibinfo{volume}{85}},
  \bibinfo{pages}{2418} (\bibinfo{year}{2000}).

\bibitem[{\citenamefont{Suzuki et~al.}(2008)\citenamefont{Suzuki, Miyakawa, and
  Suzuki}}]{suzuki2008}
\bibinfo{author}{\bibfnamefont{K.}~\bibnamefont{Suzuki}},
  \bibinfo{author}{\bibfnamefont{T.}~\bibnamefont{Miyakawa}}, \bibnamefont{and}
  \bibinfo{author}{\bibfnamefont{T.}~\bibnamefont{Suzuki}},
  \bibinfo{journal}{Phys. Rev. A} \textbf{\bibinfo{volume}{77}},
  \bibinfo{eid}{043629} (\bibinfo{year}{2008}).

\bibitem[{\citenamefont{Kalas et~al.}()\citenamefont{Kalas, Balatsky, and
  Mozyrsky}}]{kalas2008}
\bibinfo{author}{\bibfnamefont{R.~M.} \bibnamefont{Kalas}},
  \bibinfo{author}{\bibfnamefont{A.~V.} \bibnamefont{Balatsky}},
  \bibnamefont{and} \bibinfo{author}{\bibfnamefont{D.}~\bibnamefont{Mozyrsky}},
  \eprint{arXiv:0806.0419}.

\bibitem[{\citenamefont{Lewenstein et~al.}(2004)\citenamefont{Lewenstein,
  Santos, Baranov, and Fehrmann}}]{lewenstein_04}
\bibinfo{author}{\bibfnamefont{M.}~\bibnamefont{Lewenstein}},
  \bibinfo{author}{\bibfnamefont{L.}~\bibnamefont{Santos}},
  \bibinfo{author}{\bibfnamefont{M.~A.} \bibnamefont{Baranov}},
  \bibnamefont{and} \bibinfo{author}{\bibfnamefont{H.}~\bibnamefont{Fehrmann}},
  \bibinfo{journal}{Phys. Rev. Lett.} \textbf{\bibinfo{volume}{92}},
  \bibinfo{eid}{050401} (\bibinfo{year}{2004}).

\bibitem[{\citenamefont{Stan et~al.}(2004)\citenamefont{Stan, Zwierlein,
  Schunck, Raupach, and Ketterle}}]{stan_04}
\bibinfo{author}{\bibfnamefont{C.~A.} \bibnamefont{Stan}},
  \bibinfo{author}{\bibfnamefont{M.~W.} \bibnamefont{Zwierlein}},
  \bibinfo{author}{\bibfnamefont{C.~H.} \bibnamefont{Schunck}},
  \bibinfo{author}{\bibfnamefont{S.~M.~F.} \bibnamefont{Raupach}},
  \bibnamefont{and} \bibinfo{author}{\bibfnamefont{W.}~\bibnamefont{Ketterle}},
  \bibinfo{journal}{Phys. Rev. Lett.} \textbf{\bibinfo{volume}{93}},
  \bibinfo{eid}{143001} (\bibinfo{year}{2004}).

\bibitem[{\citenamefont{Inouye et~al.}(2004)\citenamefont{Inouye, Goldwin,
  Olsen, Ticknor, Bohn, and Jin}}]{inouye_04}
\bibinfo{author}{\bibfnamefont{S.}~\bibnamefont{Inouye}},
  \bibinfo{author}{\bibfnamefont{J.}~\bibnamefont{Goldwin}},
  \bibinfo{author}{\bibfnamefont{M.~L.} \bibnamefont{Olsen}},
  \bibinfo{author}{\bibfnamefont{C.}~\bibnamefont{Ticknor}},
  \bibinfo{author}{\bibfnamefont{J.~L.} \bibnamefont{Bohn}}, \bibnamefont{and}
  \bibinfo{author}{\bibfnamefont{D.~S.} \bibnamefont{Jin}},
  \bibinfo{journal}{Phys. Rev. Lett.} \textbf{\bibinfo{volume}{93}},
  \bibinfo{eid}{183201} (\bibinfo{year}{2004}).

\bibitem[{\citenamefont{Ferlaino et~al.}(2006)\citenamefont{Ferlaino, D'Errico,
  Roati, Zaccanti, Inguscio, Modugno, and Simoni}}]{ferlaino_06}
\bibinfo{author}{\bibfnamefont{F.}~\bibnamefont{Ferlaino}},
  \bibinfo{author}{\bibfnamefont{C.}~\bibnamefont{D'Errico}},
  \bibinfo{author}{\bibfnamefont{G.}~\bibnamefont{Roati}},
  \bibinfo{author}{\bibfnamefont{M.}~\bibnamefont{Zaccanti}},
  \bibinfo{author}{\bibfnamefont{M.}~\bibnamefont{Inguscio}},
  \bibinfo{author}{\bibfnamefont{G.}~\bibnamefont{Modugno}}, \bibnamefont{and}
  \bibinfo{author}{\bibfnamefont{A.}~\bibnamefont{Simoni}},
  \bibinfo{journal}{Phys. Rev. A} \textbf{\bibinfo{volume}{73}},
  \bibinfo{pages}{040702(R)} (\bibinfo{year}{2006}).

\bibitem[{\citenamefont{Deh et~al.}(2008)\citenamefont{Deh, Marzok, Zimmermann,
  and Courteille}}]{deh2008}
\bibinfo{author}{\bibfnamefont{B.}~\bibnamefont{Deh}},
  \bibinfo{author}{\bibfnamefont{C.}~\bibnamefont{Marzok}},
  \bibinfo{author}{\bibfnamefont{C.}~\bibnamefont{Zimmermann}},
  \bibnamefont{and} \bibinfo{author}{\bibfnamefont{P.~W.}
  \bibnamefont{Courteille}}, \bibinfo{journal}{Phys. Rev. A}
  \textbf{\bibinfo{volume}{77}}, \bibinfo{eid}{010701(R)} (\bibinfo{year}{2008}).

\bibitem[{\citenamefont{Ospelkaus
  et~al.}(2006{\natexlab{a}})\citenamefont{Ospelkaus, Ospelkaus, Sengstock, and
  Bongs}}]{ospelkaus_06_3}
\bibinfo{author}{\bibfnamefont{C.}~\bibnamefont{Ospelkaus}},
  \bibinfo{author}{\bibfnamefont{S.}~\bibnamefont{Ospelkaus}},
  \bibinfo{author}{\bibfnamefont{K.}~\bibnamefont{Sengstock}},
  \bibnamefont{and} \bibinfo{author}{\bibfnamefont{K.}~\bibnamefont{Bongs}},
  \bibinfo{journal}{Phys. Rev. Lett.} \textbf{\bibinfo{volume}{96}},
  \bibinfo{pages}{020401} (\bibinfo{year}{2006}{\natexlab{a}}).

\bibitem[{\citenamefont{Ospelkaus
  et~al.}(2006{\natexlab{b}})\citenamefont{Ospelkaus, Ospelkaus, Humbert,
  Sengstock, and Bongs}}]{ospelkaus_06}
\bibinfo{author}{\bibfnamefont{S.}~\bibnamefont{Ospelkaus}},
  \bibinfo{author}{\bibfnamefont{C.}~\bibnamefont{Ospelkaus}},
  \bibinfo{author}{\bibfnamefont{L.}~\bibnamefont{Humbert}},
  \bibinfo{author}{\bibfnamefont{K.}~\bibnamefont{Sengstock}},
  \bibnamefont{and} \bibinfo{author}{\bibfnamefont{K.}~\bibnamefont{Bongs}},
  \bibinfo{journal}{Phys. Rev. Lett.} \textbf{\bibinfo{volume}{97}},
  \bibinfo{pages}{120403} (\bibinfo{year}{2006}{\natexlab{b}}).

\bibitem[{\citenamefont{Zaccanti et~al.}(2006)\citenamefont{Zaccanti, D'Errico,
  Ferlaino, Roati, Inguscio, and Modugno}}]{zaccanti_06}
\bibinfo{author}{\bibfnamefont{M.}~\bibnamefont{Zaccanti}},
  \bibinfo{author}{\bibfnamefont{C.}~\bibnamefont{D'Errico}},
  \bibinfo{author}{\bibfnamefont{F.}~\bibnamefont{Ferlaino}},
  \bibinfo{author}{\bibfnamefont{G.}~\bibnamefont{Roati}},
  \bibinfo{author}{\bibfnamefont{M.}~\bibnamefont{Inguscio}}, \bibnamefont{and}
  \bibinfo{author}{\bibfnamefont{G.}~\bibnamefont{Modugno}},
  \bibinfo{journal}{Phys. Rev. A} \textbf{\bibinfo{volume}{74}},
  \bibinfo{pages}{041605(R)} (\bibinfo{year}{2006}).

\bibitem[{\citenamefont{Modugno}()}]{modugno_07}
\bibinfo{author}{\bibfnamefont{G.}~\bibnamefont{Modugno}},
  \eprint{cond-mat/0702277}.

\bibitem[{\citenamefont{Zirbel et~al.}(2008)\citenamefont{Zirbel, Ni,
  Ospelkaus, D'Incao, Wieman, Ye, and Jin}}]{zirbel07}
\bibinfo{author}{\bibfnamefont{J.~J.} \bibnamefont{Zirbel}},
  \bibinfo{author}{\bibfnamefont{K.-K.} \bibnamefont{Ni}},
  \bibinfo{author}{\bibfnamefont{S.}~\bibnamefont{Ospelkaus}},
  \bibinfo{author}{\bibfnamefont{J.~P.} \bibnamefont{D'Incao}},
  \bibinfo{author}{\bibfnamefont{C.~E.} \bibnamefont{Wieman}},
  \bibinfo{author}{\bibfnamefont{J.}~\bibnamefont{Ye}}, \bibnamefont{and}
  \bibinfo{author}{\bibfnamefont{D.~S.} \bibnamefont{Jin}},
  \bibinfo{journal}{Phys. Rev. Lett.} \textbf{\bibinfo{volume}{100}},
  \bibinfo{eid}{143201} (\bibinfo{year}{2008}).

\bibitem[{\citenamefont{Adhikari}(2004)}]{adhikari2004}
\bibinfo{author}{\bibfnamefont{S.~K.} \bibnamefont{Adhikari}},
  \bibinfo{journal}{Phys. Rev. A} \textbf{\bibinfo{volume}{70}},
  \bibinfo{pages}{043617} (\bibinfo{year}{2004}).

\bibitem[{\citenamefont{Modugno et~al.}(2003)\citenamefont{Modugno, Ferlaino,
  Riboli, Roati, Modugno, and Inguscio}}]{modugno03}
\bibinfo{author}{\bibfnamefont{M.}~\bibnamefont{Modugno}},
  \bibinfo{author}{\bibfnamefont{F.}~\bibnamefont{Ferlaino}},
  \bibinfo{author}{\bibfnamefont{F.}~\bibnamefont{Riboli}},
  \bibinfo{author}{\bibfnamefont{G.}~\bibnamefont{Roati}},
  \bibinfo{author}{\bibfnamefont{G.}~\bibnamefont{Modugno}}, \bibnamefont{and}
  \bibinfo{author}{\bibfnamefont{M.}~\bibnamefont{Inguscio}},
  \bibinfo{journal}{Phys. Rev. A} \textbf{\bibinfo{volume}{68}},
  \bibinfo{pages}{043626} (\bibinfo{year}{2003}).

\bibitem[{\citenamefont{Chui et~al.}(2004)\citenamefont{Chui, Ryzhov, and
  Tareyeva}}]{chui04}
\bibinfo{author}{\bibfnamefont{S.~T.} \bibnamefont{Chui}},
  \bibinfo{author}{\bibfnamefont{V.~N.} \bibnamefont{Ryzhov}},
  \bibnamefont{and} \bibinfo{author}{\bibfnamefont{E.~E.}
  \bibnamefont{Tareyeva}}, \bibinfo{journal}{JETP Lett.}
  \textbf{\bibinfo{volume}{80}}, \bibinfo{pages}{274} (\bibinfo{year}{2004}).

\bibitem[{\citenamefont{Graf et~al.}(1967)\citenamefont{Graf, Lee, and
  Reppy}}]{graf67}
\bibinfo{author}{\bibfnamefont{E.~H.} \bibnamefont{Graf}},
  \bibinfo{author}{\bibfnamefont{D.~M.} \bibnamefont{Lee}}, \bibnamefont{and}
  \bibinfo{author}{\bibfnamefont{J.~D.} \bibnamefont{Reppy}},
  \bibinfo{journal}{Phys. Rev. Lett.} \textbf{\bibinfo{volume}{19}},
  \bibinfo{pages}{417} (\bibinfo{year}{1967}).

\bibitem[{\citenamefont{Andreev et~al.}(2004)\citenamefont{Andreev, Gurarie,
  and Radzihovsky}}]{andreev2004}
\bibinfo{author}{\bibfnamefont{A.~V.} \bibnamefont{Andreev}},
  \bibinfo{author}{\bibfnamefont{V.}~\bibnamefont{Gurarie}}, \bibnamefont{and}
  \bibinfo{author}{\bibfnamefont{L.}~\bibnamefont{Radzihovsky}},
  \bibinfo{journal}{Phys. Rev. Lett.} \textbf{\bibinfo{volume}{93}},
  \bibinfo{pages}{130402} (\bibinfo{year}{2004}).

\bibitem[{\citenamefont{Parish et~al.}(2007)\citenamefont{Parish, Marchetti,
  Lamacraft, and Simons}}]{parish2007}
\bibinfo{author}{\bibfnamefont{M.~M.} \bibnamefont{Parish}},
  \bibinfo{author}{\bibfnamefont{F.~M.} \bibnamefont{Marchetti}},
  \bibinfo{author}{\bibfnamefont{A.}~\bibnamefont{Lamacraft}},
  \bibnamefont{and} \bibinfo{author}{\bibfnamefont{B.~D.}
  \bibnamefont{Simons}}, \bibinfo{journal}{Nature Phys.}
  \textbf{\bibinfo{volume}{3}}, \bibinfo{pages}{124} (\bibinfo{year}{2007}).

\bibitem[{\citenamefont{Kagan et~al.}(2004)\citenamefont{Kagan, Brodsky,
  Efremov, and Klaptsov}}]{kagan04}
\bibinfo{author}{\bibfnamefont{M.~Y.} \bibnamefont{Kagan}},
  \bibinfo{author}{\bibfnamefont{I.~V.} \bibnamefont{Brodsky}},
  \bibinfo{author}{\bibfnamefont{D.~V.} \bibnamefont{Efremov}},
  \bibnamefont{and} \bibinfo{author}{\bibfnamefont{A.~V.}
  \bibnamefont{Klaptsov}}, \bibinfo{journal}{Phys. Rev. A}
  \textbf{\bibinfo{volume}{70}}, \bibinfo{pages}{023607}
  (\bibinfo{year}{2004}).

\bibitem[{\citenamefont{Storozhenko et~al.}(2005)\citenamefont{Storozhenko,
  Schuck, Suzuki, Yabu, and Dukelsky}}]{storozhenko05}
\bibinfo{author}{\bibfnamefont{A.}~\bibnamefont{Storozhenko}},
  \bibinfo{author}{\bibfnamefont{P.}~\bibnamefont{Schuck}},
  \bibinfo{author}{\bibfnamefont{T.}~\bibnamefont{Suzuki}},
  \bibinfo{author}{\bibfnamefont{H.}~\bibnamefont{Yabu}}, \bibnamefont{and}
  \bibinfo{author}{\bibfnamefont{J.}~\bibnamefont{Dukelsky}},
  \bibinfo{journal}{Phys. Rev. A} \textbf{\bibinfo{volume}{71}},
  \bibinfo{pages}{063617} (\bibinfo{year}{2005}).

\bibitem[{\citenamefont{Gacesa et~al.}(2008)\citenamefont{Gacesa, Pellegrini,
  and C\^{o}t\'{e}}}]{gacesa2008}
\bibinfo{author}{\bibfnamefont{M.}~\bibnamefont{Gacesa}},
  \bibinfo{author}{\bibfnamefont{P.}~\bibnamefont{Pellegrini}},
  \bibnamefont{and}
  \bibinfo{author}{\bibfnamefont{R.}~\bibnamefont{C\^{o}t\'{e}}},
  \bibinfo{journal}{Phys. Rev. A} \textbf{\bibinfo{volume}{78}},
  \bibinfo{eid}{010701(R)} (\bibinfo{year}{2008}).

\bibitem[{\citenamefont{Yabu et~al.}(2003)\citenamefont{Yabu, Takayama, and
  Suzuki}}]{yabu2003}
\bibinfo{author}{\bibfnamefont{H.}~\bibnamefont{Yabu}},
  \bibinfo{author}{\bibfnamefont{Y.}~\bibnamefont{Takayama}}, \bibnamefont{and}
  \bibinfo{author}{\bibfnamefont{T.}~\bibnamefont{Suzuki}},
  \bibinfo{journal}{Physica B} \textbf{\bibinfo{volume}{329-333}},
  \bibinfo{pages}{25} (\bibinfo{year}{2003}).

\bibitem[{\citenamefont{Lamacraft and Marchetti}(2008)}]{lamacraft07}
\bibinfo{author}{\bibfnamefont{A.}~\bibnamefont{Lamacraft}} \bibnamefont{and}
  \bibinfo{author}{\bibfnamefont{F.~M.} \bibnamefont{Marchetti}},
  \bibinfo{journal}{Phys. Rev. B} \textbf{\bibinfo{volume}{77}},
  \bibinfo{eid}{014511} (\bibinfo{year}{2008}).

\bibitem[{\citenamefont{D'Incao and Esry}(2005)}]{dincao2005}
\bibinfo{author}{\bibfnamefont{J.~P.} \bibnamefont{D'Incao}} \bibnamefont{and}
  \bibinfo{author}{\bibfnamefont{B.~D.} \bibnamefont{Esry}},
  \bibinfo{journal}{Phys. Rev. Lett.} \textbf{\bibinfo{volume}{94}},
  \bibinfo{pages}{213201} (\bibinfo{year}{2005}).

\bibitem[{\citenamefont{Petrov}(2004)}]{petrov2004}
\bibinfo{author}{\bibfnamefont{D.~S.} \bibnamefont{Petrov}},
  \bibinfo{journal}{Phys. Rev. Lett.} \textbf{\bibinfo{volume}{93}},
  \bibinfo{pages}{143201} (\bibinfo{year}{2004}).

\bibitem[{\citenamefont{Ospelkaus
  et~al.}(2006{\natexlab{c}})\citenamefont{Ospelkaus, Ospelkaus, Humbert,
  Ernst, Sengstock, and Bongs}}]{ospelkaus_06_2}
\bibinfo{author}{\bibfnamefont{C.}~\bibnamefont{Ospelkaus}},
  \bibinfo{author}{\bibfnamefont{S.}~\bibnamefont{Ospelkaus}},
  \bibinfo{author}{\bibfnamefont{L.}~\bibnamefont{Humbert}},
  \bibinfo{author}{\bibfnamefont{P.}~\bibnamefont{Ernst}},
  \bibinfo{author}{\bibfnamefont{K.}~\bibnamefont{Sengstock}},
  \bibnamefont{and} \bibinfo{author}{\bibfnamefont{K.}~\bibnamefont{Bongs}},
  \bibinfo{journal}{Phys. Rev. Lett.} \textbf{\bibinfo{volume}{97}},
  \bibinfo{pages}{120402} (\bibinfo{year}{2006}{\natexlab{c}}).

\end{thebibliography}

\end{document}